\documentclass[preprint]{aastex}
\newcommand{\gsim}{\mbox{$\stackrel {>}{_{\sim}}$}}
\newcommand{\lsim}{\mbox{$\stackrel {<}{_{\sim}}$}}

\slugcomment{submitted to the Astronomical Journal}

\shorttitle{CO in NGC 3077}
\shortauthors{Meier et al.}

\begin{document}

\title{Molecular Gas and Star Formation in NGC 3077}

\author{David S. Meier and Jean L. Turner} 
\affil{Department of Physics and Astronomy,
University of California, Los Angeles, CA 90095--1562
\\email: meierd;turner@astro.ucla.edu}
\and
\author{Sara C. Beck}
\affil{Department of Physics and Astronomy, Tel Aviv University,
Ramat Aviv, Israel
\\email:sara@wise.tau.ac.il}

\begin{abstract}

We present high resolution ($\sim2.^{''}5$) CO(1-0) and CO(2-1) images
of the central kiloparsec of NGC 3077 made with the Owens Valley
Millimeter Array.  CO emission is distributed in three major complexes
which resolve into at least seven GMCs.  Two complexes are associated
with the central starburst.  A third, more distant complex is not
associated with strong star formation.  The GMCs are $\sim$ 70 pc
in size and contain $\sim 10^{6}~M_{\odot}$ of molecular gas.  The
Galactic conversion factor appears applicable to NGC 3077, consistent
with its solar metallicity.  Galactic rotation in NGC 3077 is detected
for the first time in the molecular gas.  The molecular gas
counterrotates with respect to the large scale HI tidal bridge.  The
molecular clouds closest to the starburst have super-virial
linewidths, possibly related to the turbulence generated by the
starburst.  The 2.6 mm radio continuum flux indicates that thermal
Bremsstrahlung dominates the emission from the starburst region below
5 cm, and that the $N_{Lyc}~\simeq ~ 3.7\times 10^{52}~s^{-1}$ or
$\sim$3000 O7 stars, corresponds to a star formation rate of 0.4
$M_{\odot}~yr^{-1}$.  At this rate the amount of molecular gas can
sustain star formation for only $\sim$10 Myrs; thus NGC 3077 is a true
starburst galaxy.  The derived age of the starburst is consistent with
the inferred ages of the superbubbles, which suggests that the burst
is much younger than the age of the M 81-M 82-NGC 3077 interaction.
We suggest that it is caused by gas that was pulled out of NGC 3077
during the interaction with M 81, raining back down onto the galaxy.

\end{abstract}
\section{Introduction}

Localized star formation in dwarf galaxies can be as intense as that
in large spiral galaxies.  Dwarf starburst galaxies appear to have
some of the most dramatic examples of ``super-star-cluster'' formation
\citep*[eg.][]{OGH94,MHLKRG95,MHST96}.  However dwarf galaxies do not
contain the usual triggering mechanisms believed to operate in large
spirals, such as spiral arms or bars.  Many young starbursts in dwarfs
(especially the galaxies characterized by Wolf-Rayet emission) seem to
have been triggered by tidal/interaction events \citep[]{B00}.
Possible perpetrators of these interaction episodes are massive
spirals, dwarfs or even intergalactic gas clouds.  The impact of these
interactions on the gas can have important consequences on the nature
and mode of star formation.  NGC 3077 is a nearby example of a 
galaxy that has experienced a strong tidal interaction in the recent 
past and is now in a starburst phase.

NGC 3077 is the somewhat overlooked third member of the widely-studied
and closest group of strongly interacting galaxies, the M 81 system
(D= 3.9 Mpc; Table \ref{tab1}).  The morphological classification of
NGC 3077 is ambiguous, perhaps due to its recent interaction with the
of the members of the group, M 81 and M 82.  Signs of interactions
between the galaxies are well established.  Large HI streamers/bridges
connect NGC 3077 to M 81.  In addition, there is an extended ridge of
atomic gas southeast of NGC 3077's nucleus with more HI than found in
the galaxy nucleus itself
\citep*[][]{C76,vH79,ADS81,YHL94,WH99,WWMS01}.  While M 82, the other
dwarf interacting with M81, has gotten most of the attention, NGC 3077
is also undergoing a burst of recent star formation
\citep*[][]{BBT74,PG89,TWK91}.  The total atomic mass content near NGC
3077 ($M_{HI}\sim 5 \times 10^{8}~M_{\odot}$) is similar to that found
in M 82 \citep*[]{C76,vH79,WWMS01,CRC78,YHL93}.  Unlike M 82 though,
the majority of the atomic gas is outside the optical confines of the
galaxy \citep[]{WH99,WWMS01}.  Moreover, the total molecular gas
content differs by nearly two orders of magnitude
\citep*[eg.][]{YS84,BSH89}.  The cause of these differences are not
well understood.  In fact, numerical models have yet to reach
consensus whether the gas stripped out of a previously gas-rich NGC
3077 or stolen from M 81 during the interaction
\citep*[][]{C76,vH79,BBCKB91,DT93,TD93,YHL94,Y98}

Lower resolution CO studies of NGC 3077 show that it contains a large
molecular cloud complex ($\sim 10^{7}~M_{\odot}$) at its center
\citep[][]{BSH89,TC92}.  It is probably this molecular cloud complex
that provides the fuel for the current starburst event.  The goal of
this paper is to map out the molecular complex at high resolution in
CO(1-0) to look for further clues on how the starburst in NGC 3077 has
proceeded, and to access the validity of the galactic conversion factor
in this metal rich dwarf.  Also simultaneous observations of CO(2-1)
are used to determine molecular gas temperatures and densities.

\section{Observations}

Simultaneous aperture synthesis observations of the CO(1-0) transition
(115.271 GHz) and the CO(2-1) transition (230.538 GHz) were made with
the Owens Valley Radio Observatory (OVRO) Millimeter Interferometer
between 1997 October 12 and 1997 December 09 (Table \ref{tab2}).  The
interferometer consists of six 10.4 m antennas with cryogenically
cooled SIS receivers \citep[][]{OVRO91,OVRO94}.  System temperatures
(single sideband) ranged from 400 - 1800 K at 115 GHz and 500 - 1100 K
at 230 GHz.  A 64 channel, 2 MHz filterbank was used to cover each
transition.  This gives a velocity resolution of 5.2 km s$^{-1}$ (2.6
km s$^{-1}$) for CO(1-0) (CO(2-1)), corresponding to an overall
bandwidth of 333 km $^{-1}$ (166 km s$^{-1}$).  The CO(2-1) data have
been smoothed to 5.2 km s$^{-1}$ to increase S/N and to match the
velocity resolution of CO(1-0).  The systemic velocity (LSR) was set
to be 7.0 km s$^{-1}$ centered at channel 32.5.  Two pointings with
phase centers of $\alpha_{1}$(B1950) = 09:59:20.0; $\delta_{1}$(B1950)
= 68:58:34.0, $\alpha_{2}$(B1950) = 09:59:18.3; $\delta_{2}$(B1950) =
68:58:40.0, were mosaicked to cover the nuclear starburst.
The data were calibrated using the MMA software package.  Phase
calibration was done by observing the quasars 1044+719 and 0923+392
every 20 minutes.  Absolute flux calibration was done using Neptune
and Uranus as primary flux calibrators and 3C273 as secondary flux
calibration.  Tracking variations in the measured flux of 3C273 with
time implies that absolute fluxes are good to 10\% at 3 mm and 20\% at
1 mm.

The two pointings were mosaicked using the MIRIAD software package.
The maps are naturally weighted and primary-beam corrected.  The OVRO
primary beam is $\sim$64$^{''}$ at (1-0) and $\sim$34$^{''}$ at (2-1).
Since the maps are mosaics, the noise level varies across the maps.
The noise increases towards the edge of the field both because of the
decreased sensitivity due to the primary beam and due to a factor of
$\sim$2 less integration time.  Reported noise levels for this paper
are those measured from line free regions of the map half-way between
the map center and the FWHM points.  The noise level is a bit lower
than this in the center ($\sim$ 10\%) and somewhat higher than this
towards the edges of the map ($\sim\sqrt{2}$).  All further data
reduction, analysis and manipulation was done using the NRAO AIPS
package.

The shortest baselines sampled in the dataset are $\simeq$15 m.  This
corresponds to spatial scales of $\sim$36$^{''}$ for (1-0) and
$\sim$18$^{''}$ for (2-1).  Structures extended on scales larger than
this will be resolved out by the interferometer.  The amount of flux
resolved out has been estimated by convolving both maps to the
resolution of the single-dish data \citep[]{BSH89}.  The peak CO(1-0)
(CO(2-1)) intensities in the OVRO maps, when convolved to a 21$^{''}$
(13$^{''}$) beamsize are 3.0 K km s$^{-1}$ (3.8 K km s$^{-1}$).  We
therefore detect $\sim$80 \% ($\sim$60 \%) of the peak intensities
obtained from the single-dish observations.

\section{Results}
\subsection{Molecular Gas Morphology}

The CO channel maps of NGC 3077 are presented in Figures 1-2.
Emission is detected in channels from V$_{LSR}$ = -30 to +30 km
s$^{-1}$ with peak antenna temperatures of $\sim$ 3.0 K for CO(1-0)
and 1.9 K for CO(2-1) (Table \ref{tab3}).  Integrated intensity maps
are plotted to the same scale in Figure 3.  The molecular gas is
concentrated into three major complexes, consistent with what is seen
in lower resolution images \citep[][]{TC92}.  The central two
complexes are extended northeast-southwest along the major axis of the
galaxy (p.a. $\sim 45^{o}$; Table \ref{tab1}).  The third complex is
found $\sim$ 550 pc in projection from the center of NGC 3077,
northwest along the minor axis.

Figure 4 shows the CO(1-0) integrated intensity map overlaid on HST
images of the galaxy in infrared and UV light
\citep[][]{MHST96,BHST99}.  Near-infrared continuum emission is likely
dominated by older stellar populations, so we use the NICMOS H band
image (Figure 4a) to locate the dynamical center of the galaxy
(assumed to be the centroid of the old stellar population).  The
brightest of the molecular gas complexes, GMCs B + D (\S 3.1.1), is
within 3$^{''}$ of the H band centroid.  The strongest star forming
region, as traced by 2.6 mm continuum (\S 3.2) and P$\alpha$ (Figure
4b), is located at the interface of these two GMCs and is labeled
``*'' in Figure 3.  Weaker star formation is seen towards the
northeastern complex, GMC A.

The correspondence between the central molecular gas distribution and
the extinction seen in the HST FOC UV image (Figure 4c) is striking.
This strongly suggests that the youngest star forming regions are
obscured by the dust associated with the molecular clouds.  The two
clusters seen in the UV (UV1 and UV2) are presumably older than the
starburst cluster, having cleared out much of their molecular gas.
The southwestern UV cluster, UV2, is found at the base of the large
superbubble seen in H$\alpha$ and P$\alpha$ \citep[eg.][Figure
4a]{M98,BHST99}, indicating that it may be the energy source of the
superbubble.  Very recently, high resolution HI observations of NGC
3077 have been presented and show that HI and the superbubbles tend to
be anti-correlated \citep[]{WWMS01}.

\subsubsection{Giant Molecular Clouds in NGC 3077}

Because of its higher sensitivity and slightly higher effective
critical density, the CO(2-1) data are used to fit the properties of
the GMCs.  Each cloud, identified as a region of spatially and
spectrally localized emission but not necessarily a gravitationally
bound entity, has been fit with an elliptical Gaussian to determine
its size, location, and intensity.  Three molecular clouds lie outside
the (2-1) primary beam.  Of these, only GMC F was fit using the
CO(1-0) data since GMC E \& G are weak or at the edge of the (1-0)
primary beam.  Once cloud sizes were determined, a box containing all
the emission from the cloud was summed to produce a spectrum.  This
spectrum was then fit with a gaussian to obtain its linewidth.

All of the fitted GMCs have been resolved except GMC C (Table
\ref{tab4}).  The three central GMCs A, B, and D are $\sim 70$ pc in
size, with implied virial masses of about $10^{6-7} ~M_{\odot}$ and
FWHM linewidths of 25 - 50 $\rm K~km~s^{-1}$ (Table \ref{tab4}).
These GMCs are similar or slightly larger in size to those in other
dwarf galaxies \citep[eg.][]{RLB93,W94,W95,THKG99,WTHSM01}, and
consistent with the larger GMCs seen in nearby spirals
\citep[eg.][]{SRBY87,VBB87,WS90,WR93,MT01}.  On the other hand, the
linewidths of the clouds in NGC 3077 are larger than seen in other
dwarf galaxies.  Typical FWHM linewidths found for $\sim$70 pc GMCs in
nearby dwarfs are $\sim$10 km s$^{-1}$
\citep[eg.][]{W94,THKG99,WTHSM01}, whereas in NGC 3077 they are 25 -
50 km s$^{-1}$.  It is possible that these GMCs are made of smaller
unresolved clumps.  In ssuch a case, the linewidths of the individual
clumps will be smaller, and our measured linewidths will reflect the
random motions of the clumps in the GMC.  Using the derived rotation
curve (\S 3.1.2), we estimate that the contribution of galactic
rotation to the linewidth is at most 5 km s$^{-1}$.  The two GMCs with
the largest linewidths are GMC B \& D.  These GMCs are the ones most
closely associated with the starburst.  GMC F, well away from the
starburst, does has a linewidth that is much smaller, and similar to
what is found for the other dwarfs.  Masses estimated assuming that
these GMCs are in virial equilibrium are compared with masses
estimated by other means in \S 4.2.  The masses estimated assuming
virialization holds tend to be larger than found by other methods.
This fact implies that at the burst GMCs additional contributions to
the linewidth besides gravitation may be present.

\subsubsection{The Velocity field}

Figure 5 displays the velocity field (moment 1) and the velocity
dispersion (moment 2) maps derived from the CO(1-0) dataset (the
CO(2-1) velocity field, not shown, is similar to CO(1-0)).  While only
a limited portion of the total HI velocity field can be traced with
CO, it appears that galactic rotation is detected for the first time
in the molecular gas.  Though small deviations are seen, the velocity
field is consistent with solid body rotation with v$_{LSR}~\simeq$ 5
km s$^{-1}$ and a total range of $\pm 30$ km s $^{-1}$ over the
central 20$^{''}$ radius along a p.a. of $\sim 45^{o}$, similar to the
optical major axis (Figure 5 insert).  Galactic rotation is $\sim1.3~
km~s^{-1}~arcsec^{-1}$, or $\sim 68 ~ km~s^{-1}~kpc^{-1}$ (corrected
for an inclination of $45^{o}$ ).

The sense of rotation of the CO velocity field is {\it opposite} of
what is seen in HI \citep[]{C76,vH79,WWMS01}.  In CO, the blueshifted
side of the galaxy is in the northeast, whereas in HI the blueshifted
side is the southwest \citep[though there some evidence for a reversal
of the HI velocity field towards the very center of the galaxy ---
Figure 12 of][]{vH79,WWMS01}.  One possible explanation for this
reversal is that the four GMCs that dominate the central emission are
rotationally decoupled from the rest of the galaxy.  A second
possibility is that the velocity field traced by the HI is not
representative of the true rotation of NGC 3077.  \citet[][]{vH79} has
suggested that the velocity field seen in the atomic gas appears to
smoothly trace that of the tidal bridge connecting NGC 3077 to M 81,
and hence the galactic rotation of NGC 3077 may not be separable from
the bulk motions of the tidal tail.  Optical spectroscopy, like CO,
suggests that the blueshifted nebular velocities are in the northeast
\citep[][]{BBT74}.  We feel that the CO velocity field probably
reflects galactic rotation and large beam HI observations do not.  The
velocity of the third major CO complex, (GMCs F \& G), located along
the minor axis, has a line center velocity of -22 km s$^{-1}$.  This
is substantially blueshifted ($\Delta v \simeq -30$ km s$^{-1}$)
relative to what would be expected from a rotating disk with the CO
velocity field.  This suggests that the cloud complex is either
falling in towards or expanding out from the nucleus.

\subsection{Millimeter Continuum Emission}

Observations of continuum emission at 2.6 mm and 1.3 mm have been made
for NGC 3077.  2.6 mm continuum is detected and Figure 6 shows the 2.6
mm continuum map generated with a 75k$\lambda$ tapered (beamsize
$3.^{''}9\times3.^{''}8, pa ~ -79^{o}$; Table \ref{tab2}).  1.3 mm
continuum emission is not convincingly detected at any location, with
a 3$\sigma$ upper limit of 9.5 mJy beam$^{-1}$.  The 2.6 mm continuum
peaks between GMCs B \& D, consistent within the uncertainties with
the dominant nebula seen in P$\alpha$.  A possible weaker source is
seen towards the UV cluster, UV2.  The peak intensity of the starburst
is 3.1$\pm$0.7 mJy beam$^{-1}$ (as measured in the $2.^{''}5$
untapered map).  The emission is slightly resolved.  The total 2.6 mm
flux summed over the central 10$^{''}$ (190 pc) centered on the
starburst cluster marked by ``*'' in Figure 3 is 8.0$\pm$2.0 mJy.  The
morphology is consistent with the higher resolution and sensitivity 20
cm VLA map of the nucleus \citep[][]{HvHKK87}.  \citet[][]{HvHKK87}
find that the central cluster has a flux of 3.6$\pm$0.3 mJy at 1.46
GHz over a deconvolved size of 11 pc x 6 pc.  The spectral index,
$\alpha$ ($S_{\nu}\propto \nu^{ \alpha}$), found for the central
cluster between 2.6 mm and 20 cm is $\alpha^{20}_{0.26}\simeq
-0.03\pm0.08$.  This flat spectral index demonstrates that the radio
emission from the central starburst HII region is dominated by thermal
Bremsstrahlung emission out to 1.4 GHz.  Single-dish observations of
the central region of NGC 3077 ($2.^{'}7$ beam at 5 cm and $1.^{'}2$
beam at 2.8 cm), obtain fluxes of 22$\pm$8 mJy at 5 cm \citep[][]{S75}
and 13$\pm$1 mJy at 2.8 cm \citep[][]{NKBW95}.  These fluxes give
spectral indices of $\alpha^{5}_{0.26}\simeq -0.34\pm0.20$ and
$\alpha^{2.8}_{0.26}\simeq -0.20\pm0.15$, respectively.  Evidently,
the radio fluxes detected in these large beams are also predominately
thermal emission and are dominated by the central 10$^{''}$ of the
galaxy.  Beyond 5 cm, synchrotron emission becomes a significant
contributor to the radio flux.

\section{Discussion}

\subsection{Physical Conditions of the Molecular Clouds in NGC 3077}

Changes in the physical conditions of the molecular gas can be traced
using the CO(2-1)/CO(1-0) ratio.  When opacities are large, the
CO(2-1)/CO(1-0) ratio is just the ratio of the Rayleigh-Jeans
corrections at the two frequencies, and hence is moderately
temperature sensitive \citep[eg.,][]{BC92}.  The physical condition of
the molecular gas in dwarfs galaxies is of interest because they may
be responsible for some portion of the observed differences in
$X_{CO}$, molecular gas abundances and star formation properties
between dwarfs and spirals \citep[][]{IDVD86,MB88,E89,S96}.  Gas
properties in dwarf galaxies such as density, temperature and opacity
appear not to differ significantly on large scales
\citep[]{SSLH92,MTCB01}, but this is not well established yet.
Physical conditions of the gas in dwarf galaxies, including NGC 3077,
have been addressed elsewhere \citep[][]{MTCB01}.  In this section we
focus on new insights gained from these higher resolution images.

The CO(2-1)/CO(1-0) ratio for both the peak antenna temperatures,
T$_{mb}^{pk}$, and integrated intensities of each GMC are displayed in
Table \ref{tab3}.  GMCs B \& D both have ratios $\sim$0.6, while GMC A
has a slightly higher ratio of $\sim$0.82.  These values are somewhat
lower than are common for starburst regions \citep[]{BC92}, but are
consistent with those of Galactic GMCs, and of nearby, less active
dwarf galaxies \citep[]{SSLH92,SHHHO97,OHHHS98,BHHFK00}.  The
T$_{mb}^{pk}$ ratios do not peak at the exact location of the
starburst in NGC 3077, unlike what is seen in the other nearby
starbursts like M 82 \citep[]{WNHK01}, IC 342 \citep[]{MT01} and
Maffei 2 \citep*[][]{MHT01}.  Ratios of 0.6 - 0.8 imply an excitation
temperature, T$_{ex}$ of 5 - 10 K for the clouds.  Comparing T$_{ex}$
to T$_{mb}^{pk}$, for each cloud provides an estimate of the GMCs
areal filling factor, $f\simeq T_{mb}^{pk}/T_{ex}$.  Typical values of
the filling factors are 0.5 for the HII region GMCs (B \& D) and 0.2
for GMC A.  Unlike the T$_{mb}^{pk}$ ratios, the CO(2-1)/CO(1-0)
integrated intensity ratios {\it do} peak towards GMC D near the
starburst site.  The difference between the integrated line ratio and
the T$_{mb}^{pk}$ ratio is explained by the CO(2-1) spectral line
being wider than the CO(1-0) line in this region.  The CO(3-2)
linewidth shows the same trend \citep[][]{MTCB01}.

Since the amount of flux resolved out by the interferometer differs
for the two transitions, the observed ratios will differ from the true
ratios by factors up to the ratio of the differences in resolved
out flux, 30\%, depending on the distribution of the resolved flux.
This effect should be less dramatic for ratios of T$_{mb}^{pk}$ since
the brightest regions of emission are generally localized in clouds.
Therefore, the ratio of T$_{mb}^{pk}$ are likely better
representations of the true ratio than ratios of integrated
intensities, but are still lower limits.  Since the amount of resolved
out flux is greater at CO(2-1) than at CO(1-0), the diffuse component
resolved out by the interferometer appears to have a ratio larger than
1, suggestive of optically thin gas.  The single-dish CO(3-2)/CO(1-0)
line ratio of 1.1 for NGC 3077 is one of the largest measured for
dwarf starbursts \citep[][]{MTCB01}, so it is not unreasonable to
expect some optically thin CO emission.

A sample of Large Velocity Gradient (LVG) radiative transfer models
were run to relate the observed T$_{mb}^{pk}$ and ratios to the
intrinsic cloud properties of density, $n_{H_{2}}$, and kinetic
temperature, T$_{k}$ \citep*[][]{GK74,SS74,DCD75}.  Details of the LVG
model used can be found in \citet{MTCB01}.  In Figure 7, two models
are displayed with $X_{CO}/dv/dr~=~ 10^{-4.1}$ and
$X_{CO}/dv/dr~=~10^{-4.5}$ for the three central GMCs.  These
abundance per velocity gradients are based on the solar neighborhood
CO/H$_{2}$ abundance \citep*[$8\times10^{-5}$;][]{FLW82} and a
velocity gradient, $dv/dr \sim 1~ km~s^{-1}$ (Table \ref{tab4}).

Best fit solutions obtained from the LVG modeling give
$<n_{H_{2}}>~\sim ~ 200~cm^{-3}$ and T$_{k}~ \gsim ~ 25$ K for
the three central GMCs.  Cooler and denser solutions are also
permitted for GMCs A and B ($<n_{H_{2}}>~\gsim ~ 10^{3}~cm^{-3}$ and
T$_{k}~ \sim ~ 10 - 20$ K).  Little difference is found between the
two solutions with different $X_{CO}/dv/dr$.  Changing the estimated
filling factors substantially from the assumed values worsen the
$\chi^{2}$ fits.  Increasing the filling factor by a factor of two
pushes the best fit solutions to slightly cooler ($\Delta$ T$_{k}~\sim
5$ K) and denser ($<\Delta n_{H_{2}}> ~\sim 0.3$ dex) solutions.
Since the ratios are lower limits due to the differences in resolved
out flux between the two transitions, the true gas densities and
temperatures are likely to be higher than the above values.
Single-dish LVG modeling using CO(3-2) data do, in fact, obtain higher
gas temperatures and densities \citep[T$_{k}~\gsim ~20$ K,
$<n_{H_{2}}>~\gsim ~ 10^{3}~cm^{-3}$;][]{MTCB01}.  Therefore, while not 
dramatic, indications of elevated kinetic temperatures are present.

\subsection{The Conversion Factor in NGC 3077}

To study the relationship between molecular gas, dynamics, and star
formation one must have an accurate accounting of the amount of
molecular gas.  Usually N$_{H_{2}}$ is estimated from I$_{CO}$ using a
conversion factor, $X_{CO}$, which is empirically determined from
Galactic molecular clouds.  There is evidence that $X_{CO}$ can vary.
In nuclear starburst regions, the intense radiation field and the
different cloud conditions due to the deep potential well can have
significant effects on $X_{CO}$ \citep[][]{MB88,E89,S96}.  While
dynamical influences on the conversion factor are expected to be less
important in dwarf galaxies, their low metallicities cause CO to be
difficult to detect \citep*{I86,ACCK88,VH95,W95,AST96}.  The fact that
NGC 3077 is a low mass dwarf with solar metallicity
\citep[][]{H80,M97,SM01} and strong star formation makes it an
interesting object in which to estimate the conversion factor.

Molecular masses estimated using the Galactic disk conversion factor,
$X_{CO_{gal}}$ = $2.3\times 10^{20}~cm^{-2}(K km s^{-1})^{-1}$
\citep{SET88,Het97}, are:
\begin{equation}
M_{mol}~=~1.23\times 10^{4}~ (M_{\odot}) D^{2}_{mpc}~S_{CO},
\end{equation}
\citep[eg.][]{WSFMS88}, where $S_{CO}$ is the CO(1-0) flux in Jy km
s$^{-1}$.  Derived masses for the GMCs using $X_{CO}$ are shown in
Table \ref{tab4}.  

To access the applicability of the Galactic conversion factor to NGC
3077, we attempt to estimate the amount of molecular gas independently
of $X_{CO}$.  Several methods exist to estimate molecular gas masses,
independent of CO(1-0).  Since the GMCs in the center of NGC 3077 are
resolved, the total mass of the GMCs can be estimated from the virial
theorem made with the assumption that they are virialized
\citep[eg.][]{MT01,MTCB01}.  The virial cloud masses assuming
$r~\propto \rho ^{\gamma}$ with $\gamma = 1$, appropriate for Galactic
clouds \citep*[eg.,][]{CBD85, SYCSW87, MRW88}, are displayed in Table
\ref{tab4}.  In general, masses obtained from the virial theorem are
larger than those obtained from the CO(1-0) conversion factor.  This
is particularly evident for GMCs B \& D, where the mass derived from
the virial theorem is an order of magnitude larger than that from the
CO conversion factor.  This stems directly from the fact that the
clouds have large and possibly nonequilibrium linewidths at these
locations (\S 3.1.2).  The virial mass of the most distant GMC, F, is
in better agreement with the conversion factor estimate there.

Optical extinction can also be used to estimate the total gas mass
present towards the starburst.  The extinction to the central
starburst is derived by comparing the H$\alpha$ flux of the starburst
to the extinction-free millimeter continuum (\S4.3), and is found to
be A$_{V}~\simeq ~2.5$.  If it is assumed that the starburst is in the
middle of the cloud, then a total extinction of 5 mag is estimated.
Assuming that N$_{H_{2}}\simeq 1\times 10^{21} ~cm^{-2}~mag^{-1}A_{V}$
\citep*[eg.][]{BSD78}, a total column density of N$_{H_{2}}\simeq
5\times 10^{21}~cm^{-2}$ is implied.  Whereas, the average CO(1-0)
intensity at the same location is $\sim 45 ~K ~km~s^{-1}$.  From these
results a conversion factor of $X_{CO}\sim 1.1\times 10^{20}~cm^{-2}
(K~km~s^{-1})^{-1}$ is estimated.  This is slightly smaller but within
a factor of two of the Galactic value.

Gas masses can also be estimated from thermal dust emission.  For a
dust temperature of 34 K \citep{MI94}, a 1.3 mm dust absorption
coefficent of $3.4 \times 10^{-3}~\rm cm^{2}~g^{-1}$ \citep{PHBSRF94}
and a gas-to-dust ratio of 100, $M_{gas}~\simeq ~ 1.15\times
10^{5}~S_{mJy}(1.3mm)~M_{\odot}$.  If we assume that all 1.3 mm flux
is due to dust emission (an upper limit since some thermal
bremsstrahlung will still contribute), the 3$\sigma$ upper limit of 10
mJy beam$^{-1}$ equates to $\sim 10^{6}M_{\odot}$.  Therefore, unless
the nature and emissivity of the dust in NGC 3077 is substantially
different from what is found for the Galaxy, the gas masses implied by
the dust emission are also significantly lower than the virial values
and more in line with the extinction values.  So we find that, in
general, the dust emission also favors the lower values of the
conversion factor.

In this context, the two GMCs (B \& D) with the largest ``virial''
masses appear not to be in virial equilibrium.  These two clouds are
the closest to the dynamical center and the starburst, and have
unusually large linewidths for their sizes (\S 3.1.1).  Additional
evidence that these clouds are not virialized can be seen by comparing
the densities implied by the virial mass with those estimated from gas
excitation (LVG modeling; Figure 7).  The dot-dashed line represents
the beam-averaged molecular cloud density implied by the virial
theorem.  In the case of GMC D, the beam-averaged value is a factor of
two larger than any acceptable parameter space allowed from the LVG
analysis and a factor of six larger than the optimal solution.  For
GMCs A \& B, there is strictly speaking some overlap with regions of
acceptable parameter space but they are again much denser than the
optimal LVG solution (\S 4.1).  If these GMCs resolve into smaller
clumps then these clumps appear to move around at velocities greater
than virialization suggests.  Detailed studies of the nearby starburst
spirals, IC 342 and Maffei 2, also show that the GMCs also have
super-virial linewidths \citep[eg.][]{MT01,MHT01}.  Significant
non-circular motions due to barred potentials are likely the cause of
the large linewidths in these two spiral galaxies, but such motions
cannot be the cause in NGC 3077.  The increased linewidths indicate
that these GMCs associated with the starburst have a higher level of
turbulence, possibly from shocks due to star formation, cloud-cloud
collisions, or the mechanical energy input from the associated
superbubbles \citep[][]{M98}.
   
In summary, the above methods imply that the CO conversion factor
applicable to NGC 3077 is approximately equal to or slightly lower
than the Galactic value of $2.3\times 10^{20}~cm^{-2}
(K~km~s^{-1})^{-1}$.  Since NGC 3077 is roughly solar metallicity,
this is reasonable.

\subsection{The Star Formation Rate in NGC 3077}

At millimeter wavelengths synchrotron emission is weak and the radio
continuum emission for normal galaxies is dominated by optically thin
free-free emission or dust emission.  Dust emission is weak at 1.3 mm
(\S 3.2), and hence should not be a significant contaminant to the 2.6
mm continuum emission.  Therefore 2.6 mm provides an excellent
estimate of the Lyman continuum flux and the strength of massive star
formation.  We use our 2.6 mm continuum detection to make an
extinction-free estimate of the strength of the starburst in NGC 3077.
The number of Lyman continuum photons implied by the 8.0 mJy of 2.6 mm
continuum is N$_{Lyc}~\simeq~3.7\pm0.8 \times 10^{52}~s^{-1}$ over the
central 10$^{''}$.  We detect N$_{Lyc}~\simeq~1.4\pm0.4 \times
10^{52}~s^{-1}$ from the central starburst cluster alone
\citep[assuming T$_{e}=10^{4}$ K, eg.][]{MH67}.  This is about six
times the strength of 30 Dor \citep[eg.][]{VRLC95}.  The number of
effective O7 stars needed to produce this N$_{Lyc}$ is $\sim$3000
(L$_{O7}~\simeq~7.4\times 10^{8} ~L_{\odot}$), of which $\sim$1000
(L$_{O7}~\simeq~2.8\times 10^{8} ~L_{\odot}$) come from the central
cluster \citep[][]{VGS96}.  Clearly, to explain the bright 2.6 mm
continuum flux, a large amount of young, massive stars must be have
been forming over the last $\sim$ 10 Myr.

N$_{Lyc}$ derived from the H$\alpha$ flux is $5.3\times10^{51}~s^{-1}$
\citep[][]{PG89,TWK91,M97}, a factor of seven lower than predicted
from the 2.6 mm data.  Extinction is almost certainly the cause of
most of the difference.  Comparison of the millimeter and the
H$\alpha$ rate implies an average extinction at H$\alpha$ of 2.1 mag,
or A$_{V}$ = 2.5 mag (assuming A$_{\lambda}~\propto~\lambda^{-1}$).
The P$\alpha$ flux measured over the same region (Figure 4b) is also
consistent with an A$_{V}$ of 2.5 - 3.0 mag \citep[][]{BHST99,WWMS01}.
The star formation rate (SFR) derived from H$\alpha$ luminosity is
$\sim 0.06 ~M_{\odot}~yr^{-1}$ \citep[eg.][]{TWK91}.  Corrected for
extinction, the implied true SFR is 0.4 $M_{\odot}~yr^{-1}$ over the
central 190 pc.  This equals the SFR predicted from the FIR Luminosity
\citep[eg.][Table \ref{tab1}]{TWK91}; the luminosity derived above
from the number of O stars also equals the observed far-infrared
luminosity.  This argues that all of the FIR luminosity comes from
light reradiated from the starburst population; other possible FIR
sources such as cirrus and dust heated by the old stellar population
do not appear to contribute significantly \citep[eg.][]{BLR89,TWK91}.

Assuming the conversion factor in NGC 3077 is similar to the solar
neighborhood (\S 4.1), there is $4.5\times 10^{6}$ M$_{\odot}$ of
molecular gas over the central 10$^{''}$ (corrected for resolved out
flux).  If the CO rotation curve accurately traces the dynamics of NGC
3077 over this region (\S 3.1.2), a dynamical mass of $\sim 9.7 \times
10^{6}$ M$_{\odot}$ obtained.  This implies a molecular mass fraction
of $\sim$50\% over the central 200 pc radius.  This large molecular
mass fraction is similar to what is found for nuclei of nearby
starburst spirals,including M 82, but it is substantially larger than
is typical for dwarf galaxies or ellipticals
\citep*{YS84,SSSS91,YD91,T94,WCH95,SDRB97}.  Even a factor of two
overestimate in the amount of molecular gas, would imply a molecular
mass fraction of $\sim$25\%, which is still high for dwarfs.

Even with a relatively large molecular mass fraction, the star
formation in NGC 3077 is unusually vigorous for its molecular mass.
\citet[][]{RY99} find $\frac{L_{H\alpha}}{M_{H2}}$ in NGC 3077 to be
one of the largest seen in nearby galaxies.  In their sample of 121
(mostly spiral) galaxies only NGC 1569 \citep[whose molecular mass is
probably underestimated due to its low metallicity;][]{THKG99} and NGC
3310 have higher values.  Given the amount of molecular gas and the
derived star formation rate, the central molecular cloud complex can
support star formation for only $\sim$ 11 Myr. The several
superbubbles in H$\alpha$ are further evidence for a shortlived
starburst \citep[][]{M98}.  The mechanical power of a wind driven
superbubble gives an age of $\lsim$10 Myr for the superbubbles,
\citet[][]{M98}, and presumably the starburst, in NGC 3077.
Therefore, it seems plausible that the current starburst epoch started
around 10 Myrs ago.

The UV clusters near GMC A have weak or no 2.6 mm continuum associated
with them and only weak P$\alpha$ emission.  These clusters that are
not embedded in molecular gas are probably somewhat older than those
seen at the interface of GMCs B \& D.  Optical/UV observation of these
clusters suggest that they are $\sim$ 10 - 100 Myr old with the older
age stars increasingly dominating as one proceeds away from the center
\citep[][]{BG81,PG89,AN00a, AN00b,SM01}.

\subsection{The Trigger of the Starburst}

In the course of an interaction redistribution of large amounts of gas
can occur, often times leading to starbursts in the host galaxies.
NGC 3077 is a dramatic example of such tidal carnage, however it is as
yet unclear what the source of the gas that has generated these tidal
features are and how they relate to the current nuclear starburst.
The gas surrounding NGC 3077 either originally belonged to gas-rich
disk associated with NGC 3077 that has been stripped out due to the
interaction \citep[][]{YHL94,Y98}, or the gas has been captured from M
81 during its most recent passage \citep{TD93,DT93}.  In either case,
the interaction has had a strong influence on the gas distribution,
and probably is responsible at some level for the starburst.

A consideration of the timescales involved provide clues as to how the
starburst and the interaction are related in detail.  Despite the
ambiguity in the source of the neutral gas surrounding NGC 3077, all
simulations of the M 81-M 82-NGC 3077 group agree that the strongest
interaction between NGC 3077 and M 81 occurred 300 - 600 Myr ago
\citep[][]{C76,vH79,BBCKB91,DT93,TD93,YHL94,Y98}.  The strength of the
starburst estimated from the 2.6 mm continuum, the superbubble sizes
\citep[][]{M98} and the UV colors \citep[][]{BG81,PG89,AN00a,
AN00b,SM01} all suggest that the current burst of star formation is
$\lsim$10 Myr old.  While the star formation of the outer parts of NGC
3077 and its halo object, ``Garland'', appear to be $\lsim$150 Myrs
old, more consistent with a direct trigger by the interaction
\citep[][]{SM01}, the central burst appears too young to have been
triggered directly.
  
Based on considerations of dynamical friction alone, it is expected
that gas should rain down on the disk of NGC 3077 for a few Gyrs after
closest interaction.  Additional evidence for infall can be seen in
our CO data: firstly, the starburst appears to be a result of the
collision of GMCs B \& D, and secondly, the complex of molecular gas
comprising GMCs E-G is located along the minor axis of NGC 3077 and is
moving with substantial net line-of-sight motion relative to the LSR
velocity of the galaxy.

To get an estimate of the relevant dynamical friction timescale for
the gas around NGC 3077, we consider the fate of the recently
discovered atomic/molecular gas cloud complex $\sim$4.8 kpc to the
southeast of NGC 3077's center \citep[][]{WH99,HW00}.
\citet[][]{WH99} suggest that this complex may be a dwarf galaxy
forming.  Whether this gas cloud (the WH complex) will form a separate
dwarf galaxy or will fall back into NGC 3077 depends on the geometry
of the system.  Despite many attempts, the exact interaction geometry
of the M 81-M 82-NGC 3077 system is still not completely understood
\citep[][]{C76,BBCKB91,DT93,TD93,YHL94,Y98}, but the models suggest
30$^{o}$ - 50$^{o}$ inclinations with respect to the plane of the sky.
Adopting an inclination of 45$^{o}$ and a total mass of NGC 3077
inside the radius of the WH complex of $8.6\times 10^{9} M_{\odot}$
\citep[][]{C76,vH79}, we estimate an escape velocity of 120 km
s$^{-1}$ at the location of WH complex.  The observed velocity of the
WH complex is $\sim$15 km s$^{-1}$, or only $\sim$10 km s$^{-1}$
different from the v$_{LSR}$ of NGC 3077, significantly less than the
escape velocity.  This implies that the WH complex must be moving
within $\sim 5^{o}$ of the plane of the sky, if it is to escape NGC
3077's gravitational potential and became a separate dwarf galaxy.

If the complex is bound to NGC 3077 it will spiral into the nucleus
due to dynamical friction.  The timescale for a cloud complex to
spiral into the center is \citep[eg., pg. 428][]{BT87,SG00}:
\begin{equation}
\tau_{df}\lsim {1.17 r_{i}^{2}v_{c} \over ln(\Lambda)G ~M}\simeq
{264~ Gyr \over ln(\Lambda)}
\left(\frac{r_{i}}{2~ kpc} \right)^{2}
\left(\frac{v_{c}}{250~ km s^{-1}} \right)
\left(\frac{10^{6}~M_{\odot}}{M} \right)
\label{df}
\end{equation} 
where $r_{i}$ is the distance between the cloud and the center of the
galaxy, $v_{c}$ is the orbital speed of the cloud, $M$ is the mass of
the cloud and $ln(\Lambda)$ is the Coulomb logarithm (the natural
logarthim of the ratio of maximum to minimum impact parameters).
There is some uncertainty in this value since the WH complex is not
likely on a circular orbit.  (Strickly speaking, the lower limit in
the eq.(2) is due to the fact that this equation applies to stellar
systems; since gas is dissipational it can lose angular momentum more
quickly.  However, internal dissipation in the cloud should not
significantly influence the derived timescale since the basic
assumption of a test mass used in this calculation should hold over
most of the orbit.  The dissipational nature of the gas should only
become important when the cloud reaches a distance from the nucleus
where tidal forces being to disrupt the cloud [$\sim 100 ~pc$ for the
WH complex]).  We adopt the following values as suitable to the WH
complex, $M\simeq 10^{8}~M_{\odot}$, $r_{i}\simeq ~4.8 ~kpc$,
v$_{c}\sim 50~ km~ s^{-1}$ \citep[][]{WH99,HW00} and $ln(\Lambda)
\simeq 5$ (where $\Lambda$ is approximated as $r_{i}$ divided by the
diameter of the cloud; also consistent with the formulation of
\citet[][]{BT87}).  For these values, the implied dynamical fiction
timescale is $\sim$600 Myrs .  This value is about the same as the
time since closest approach.  Therefore it seems reasonable that large
clouds of atomic or molecular gas similar in mass to the WH complex
have spiraled into the center of NGC 3077 with time delays consistent
with the current burst.

It is also possible that if NGC 3077 had a large disk of gas previous
to the interaction, it could have been driven into the nucleus to
trigger the burst.  Models suggest that strong tidal interactions can
set up bar modes that drive gas to the galaxy center
\citep[eg.][]{N87,BH92}.  However, this scenario is less favorable for
two reasons: (1) the timescales for gas inflow are extremely fast
\citep[a few dynamical times or less, $\sim 10^{7}$ Myr;][]{H89}, and
again would seem to trigger a burst too quickly, (2) There is no
indication of a stellar bar, or any strongly disturbed stellar
component remaining in NGC 3077.  Therefore, we favor recent infall of
gas onto the disk of the NGC 3077 as the cause of the current burst.
If NGC 3077 did have a large disk of gas as suggest by Yun et al
\citep[]{YHL94,Y98}, the above points suggest that a burst of star
formation starting a few hundred Myrs ago could have been driven.
There are indications for this somewhat older stellar population
\citep[]{AN00a,AN00b,SM01}.  This might provide hints that the gas
around NGC 3077 has been stripped from NGC 3077 itself, however this
cannot be taken as proof.  While important, how the gas arrived
outside the optical disk in the first place does not effect the
conclusion that the gas is dynamically unstable and is prone to infall
on timescales of hundreds of Myrs.  It seems quite reasonable that gas
clouds raing down onto the disk of NGC 3077 will drive starburst
eposides for the next few $\sim$ Gyrs.

There are other examples of infall triggered starbursts, such as the
Galactic disk \citep[the high velocity clouds, eg.][]{WvW97}, and
possibly nearby dwarf starbursts such as NGC 5253
\citep[][]{TBH97,MTB98}.  Like NGC 3077, NGC 5253 is much too far from
its putative partner (M 83) to have interacted in the last few Myrs
(the age of its starburst).  If these starbursts are triggered by the
gas accreting back onto the disk from a past interaction, then
significant delays can easily be explained.  If such a scenario is
correct then a dwarf galaxy may not always need a close companion in
order for its star formation to be triggered by an interaction event.
Smaller clouds of gas ($10^{6-8}~ M_{\odot}$) falling back to the
galaxy can sustain bursts of star formation for several Gyrs after
closest approach.

\section{Summary and Conclusions}

Images of NGC 3077 in the lines of CO(2-1) and CO(1-0), obtained from
the Owens Valley Millimeter Array, show CO emission in three major
complexes.  Two of the complexes are associated with the central
starburst and the third is along the minor axis at a distance of
$\sim$0.5 kpc.  The three complexes resolve into at least seven GMCs,
with diameters of $\sim$ 70 pc and masses of $\sim 10^{6}~M_{\odot}$,
similar to large GMCs in the Galaxy and in other nearby gas-rich
dwarfs.  Linewidths of the two GMCs closest to the starburst are quite
large, suggesting that either winds, outflows or accretion flows are
important in these clouds.  Galactic rotation is detected for the
first time in the molecular gas.  The molecular gas is counterrotating
with respect to the large scale HI tidal bridge.  Molecular gas makes
up $\sim$50 \% of the dynamical mass in the central 10$^{''}$.

The CO(2-1)/CO(1-0) antenna temperature ratio in NGC 3077 ranges from
0.6 - 0.82, corresponding to T$_{ex}~\simeq$ 5 - 10 K, typical of
Galactic GMCs.  Densities and kinetic temperatures obtained from LVG
modeling indicate that the central GMCs have $<n_{H_{2}}>~\gsim ~
200~cm^{-3}$ and T$_{k}~ \gsim ~ 25$ K.

Three methods are compared to estimate the mass of molecular hydrogen
and hence assess the validity of the Galactic conversion factor is NGC
3077.  The 1.3 mm dust continuum emission upper limits and the optical
dust extinction methods both yield masses consistent with estimates
based on the Galactic conversion factor.  Virial masses are slightly
higher but consistent with the CO masses, except for the two GMCs
closest to the starburst, which are probably not in virial
equilibrium. The Galactic conversion factor appears applicable to NGC
3077, consistent with its solar metallicity.

2.6 mm continuum emission is also detected coincident with the main
P$\alpha$ nebula, between the two largest GMCs.  The spectrum is flat
from 2.6 mm to 5 cm establishing that the bulk of the radio continuum
emission is thermal free-free emission.  The strength of the continuum
implies that the central starburst region has $N_{Lyc}~\simeq ~
3.7\times 10^{52}~s^{-1}$ or $\sim$3000 O7 stars, 5 - 10 times that
predicted from H$\alpha$ emission.  Based on the continuum data an
extinction-free star formation rate of 0.4 $M_{\odot}~yr^{-1}$ is
derived for the central 190 pc of NGC 3077.  At this rate the amount
of molecular gas present can sustain star formation for only $\sim$10
Myrs.  This age is consistent with the age of the superbubbles.  We
suggest that the burst is much younger than the age of the M 81-NGC
3077 interaction.  (1) The youth of the burst, (2) the location of the
burst at the interface between two GMCs, (3) the presence of molecular
clouds distributed along the minor axis, and (4) the large amounts of
atomic and molecular gas just outside the optical galaxy all lead to
the conclusion that the current burst resulted from neutral gas that
was torn out of the galaxy, raining back down on the disk.  Dynamical
considerations suggest that recently discovered molecular complex to
the southeast of the galaxy is bound will also fall back into NGC 3077
rather than become a new dwarf galaxy.

\acknowledgements

We are grateful to the faculty, staff and postdocs at OVRO for their
support and assistance during the observations.  We thank an anonymous
referee for providing comments that help improve the paper.  This work
is supported in part by NSF grant AST-0071276.  The Owens Valley
Millimeter Interferometer is operated by Caltech with support from the
NSF.

\clearpage

\figcaption[f1.ps]{CO(1-0) channel map.  LSR velocities
are listed at the top right of each plane.  The contours are plotted
in intervals of 55 mJy bm$^{-1}$ (or 0.75 K for the 2.$^{''}$9 x
2.$^{''}$3 beam), corresponding to 2$\sigma$.  The beam is plotted in
the bottom left of the first plane.}

\figcaption[f2.ps]{CO(2-1) channel map.  LSR velocities are
listed at the top right of each plane.  The contours are plotted in
intervals of 0.11 Jy bm$^{-1}$ (0.38 K) corresponding to 2$\sigma$.
The beamsize has been smoothed to the same resolution as the CO(1-0)
data.}

\figcaption[f3.ps]{The integrated intensity of the two
transitions. (a) CO(1-0) integrated intensity.  The contours are in
steps of the theoretical 2$\sigma$ value of 8.9 K km s$^{-1}$. The
fitted positions of the GMCs are labeled on plot.  The location of the
starburst is marked by the label, ``S'', and the positions of the two
UV clusters are also labeled. (b) CO(2-1) integrated intensity.
Contours are in steps of the 3$\sigma$ value of 8.6 K km s$^{-1}$.
The large oval represents the FWHM power points of the mosaicked (2-1)
field.  The true map noise is slightly lower than the above quoted
values since the channel maps were clipped at 1.2$\sigma$ when making
the integrated intensity maps.  }

\figcaption[f4.eps]{The CO(1-0) integrated intensity overlaid
on the HST images.  (a) The CO(1-0) overlaid on the HST NICMOS F160W
(H band) continuum image \citep[]{BHST99}.  The contours are the same
as in Figure 3. (b) The CO(1-0) overlaid on the HST NICMOS F187N
(P$\alpha$) image \citep[]{BHST99}.  For reference, the white asterisk
marks the fitted cetroid of the 2.6 mm continuum.  (c) The CO(1-0)
overlaid on the HST FOC F220W UV image \citep[]{MHST96}.}

\figcaption[f5.ps]{The velocity field as measured from the
CO(1-0) data. (a) The velocity field (moment 1 map) of NGC 3077.
Contours are in steps of 5 km s$^{-1}$, with dashed contours
representing negative velocities.  The greyscale runs from -27 km
s$^{-1}$ to +27 km s$^{-1}$.  The bold dashed line marks the major
axis of the galaxy.  The two bold ticks mark the 0 km s$^{-1}$
contour.  The insert displays the rotation curve taken along the major
axis (dotted line).  (b) The velocity dispersion (moment 2) map.
Contours are in increments of 5 km s$^{-1}$.  The greyscale runs from
0 km s$^{-1}$ to +21 km s$^{-1}$.}

\figcaption[f6.ps]{The 2.6 mm continuum image.  The 112 GHz
continuum map displayed with a 75k$\lambda$ taper.  Contours are in
steps of the 2$\sigma$ value of 2.5 mJy beam$^{-1}$.  Dashed contours
represent negative values.  The beam is plotted in the lower left.}

\figcaption[f7.ps]{The LVG modeling for (a) GMC D, (b) GMC B, and (c)
GMC A of NGC3077.  Contoured regions represent 1$\sigma$ confidence
solutions for the observed CO(1-0) antenna temperature and the
CO(2-1)/CO(1-0) temperature ratio.  Asterisks mark the location of the
best fit (minimum $\chi^{2}$).  Gray lines are the solutions for a
representative value of $X_{12CO}/dv/dr ~ =~ 10^{-4.1}$, which
corresponds to a solar [CO/H$_{2}$] abundance of $8\times 10^{-5}$ and
a velocity gradient of 1 $km~s^{-1}~pc^{-1}$, consistent with the
data, Table \ref{tab4}.  Black lines represent a lower metallicity /
higher velocity gradient model with $X_{12CO}/dv/dr ~ =~ 10^{-4.5}$
for comparison.  The fitted models assume a filling factor of 0.5 for
GMCs B \& D and a filling factor of 0.2 for GMC A (see text).
Increasing the filling factor to unity will push the best fit
solutions towards slightly lower kinetic temperatures and slightly
higher densities.  The dot-dashed line shows the beam-averaged
densities implied if each GMC is in virial equilibrium (Table
\ref{tab4}).}

\clearpage

\begin{deluxetable}{lcc}
\tablenum{1}
\tablewidth{0pt}
\tablecaption{NGC 3077  Basic Data\label{tab1}}
\tablehead{\colhead{Characteristic} 	& \colhead{Value}	&
\colhead{Reference}}
\startdata
Revised Hubble Class      &I0pec   & 1  \\
Dyn. Center     & $\alpha(2000) = 10^{h} 03^{m} 19.^{s}16$ &2   \\
  &$ \delta(2000) = +68^{o} 44' 01''.4$       &   \\
V$_{lsr}$   &5.0 km s$^{-1}$   &2   \\
Adopted Distance  & 3.9 Mpc  &3   \\
Major Axis P.A. &45$^{o}$   &4   \\
L$_{IR}$ & $7.2\times 10^{8}~M_{\odot}$\tablenotemark{a} & 4 \\
M$_{HI}$ & $3.6\times 10^{8}~M_{\odot}$\tablenotemark{a,b}& 5 \\
N$_{Lyc}$(2.6 mm) & $3.7\times 10^{52}~s^{-1}$ & 2 \\
SFR(H$\alpha$) & 0.06 $M_{\odot}~yr^{-1}$\tablenotemark{a} &6,7 \\
SFR(Ext. Corrected) & 0.4 $M_{\odot}~yr^{-1}$\tablenotemark{a} &2 \\
\tablenotetext{a}{Corrected for the adopted distance}
\tablenotetext{b}{Inside the Holmberg Radius}
\tablecomments{Units of right ascension are hours, minutes and seconds, and 
units of declination are degrees, arcminutes, and arcseconds.}
\tablerefs{(1) de Vaucouleurs, de Vaucouleurs \& Corwin 1976; (2) this 
paper; (3) Sakai \& Madore 2001; (4) Melisse \& Israel 1994; 
(5) van der Hulst 1979; (6) Becker et al. 1989; (7) Thronson et al. 1991} 
\enddata
\end{deluxetable}

\clearpage

\begin{deluxetable}{lccccc}
\tablenum{2}
\tablewidth{0pt}
\tablecaption{Observational Data\tablenotemark{a}\label{tab2}}
\tablehead{
\colhead{Transition\tablenotemark{b}} 
&\colhead{Frequency} 
&\colhead{$\Delta V_{chan}$}
&\colhead{$\Delta \nu_{band}$} 
& \colhead{Beamsize} 
& \colhead{Noise level} \\
\colhead{} 
&\colhead{\it (GHz)} 
&\colhead{($km~s^{-1}$)}
&\colhead{\it (MHz)}
& \colhead{\it (arcsec; deg)} 
& \colhead{\it (mK / mJy Bm$^{-1}$)}} 
\startdata
CO(1-0)& 115.271 & 5.21& 128 &$2.9\times 2.3;-68^{o}$ & 380/28\\
CO(2-1)& 230.538 & 5.20\tablenotemark{c}& 128 &$2.9\times 2.3;-68^{o}$
\tablenotemark{d} & 190/55\\
3mm Cont.\tablenotemark{e} & 112.1  &\nodata  &1000 
&$3.0\times 2.3;-81^{o}$ & 6.9 /0.7\\
1mm Cont. & 230.7  &\nodata  &1000 &$3.0\times 2.3;-81^{o}$
\tablenotemark{d} & 8.3/2.5\\
\enddata
\tablenotetext{a}{For observations made from 1997 October 12 - 1997 
December 09}
\tablenotetext{b}{Phase Center \#1: $\alpha = 10^{h} 03^{m} 19^{s}.23~~
\delta = +68^{o} 44' 03.^{''}4$ (J2000)\\
$~~~~~$Phase Center \#2: $\alpha = 10^{h} 03^{m} 17^{s}.56~~
\delta = +68^{o} 44' 09.^{''}4$ (J2000)}
\tablenotetext{c}{The CO(2-1) maps were smoothed to 5.2 km s$^{-1}$ 
resolution from 2.6 km s$^{-1}$.}
\tablenotetext{d}{The 1 mm maps are smoothed to the resolution of the 
3 mm data.}
\tablenotetext{e}{For the map displayed in Figure 6, a 75k$\lambda$ 
taper is applied giving a beamsize of $3.^{''}9\times 3.^{''}8; -79^{o}$}
\end{deluxetable}

\clearpage

\begin{deluxetable}{lcccccc}
\tablenum{3} 
\tablewidth{0pt} 
\tablecaption{Measured Intensities \& Ratios\label{tab3}} 
\tablehead{ 
\colhead{GMC} 
&\colhead{T$_{mb}$(1-0)\tablenotemark{a}}
& \colhead{I$_{CO10}$\tablenotemark{a}} 
&\colhead{T$_{mb}$(2-1)\tablenotemark{a}}
& \colhead{I$_{CO21}$\tablenotemark{a}} 
& \colhead{$\frac{I_{CO(2-1)}}{I_{CO(1-0)}}$\tablenotemark{b}} 
& \colhead{$\frac{T_{CO(2-1)}}{T_{CO(1-0)}}$\tablenotemark{b}} 
\\
\colhead{}
& \colhead{K}
& \colhead{(Kkms$^{-1}$)}  
& \colhead{K}
& \colhead{(Kkms$^{-1}$)}
& \colhead{}
& \colhead{}}
\startdata
A &1.7&34&1.4&20&0.59$\pm$0.15&0.82$\pm$0.23 \\
B &2.0&57&1.2&28&0.49$\pm$0.12&0.60$\pm$0.16 \\
C &3.0&51&1.9&45&0.88$\pm$0.20&0.63$\pm$0.11 \\
D &1.1&18&0.5&9&0.50$\pm$0.22&0.45$\pm$0.24 \\
E &1.1&19&0.8&16&0.84$\pm$0.30&0.73$\pm$0.38 \\
F &3.0&27&\nodata    & \nodata&\nodata      &\nodata       \\
G &2.6&29&\nodata    & \nodata&\nodata      &\nodata       \\
\tablenotetext{a}{Uncertainties are based on the larger of the statistical 
noise or the absolute calibration uncertainty, and are $\simeq$10\% for 
(1-0) and $\simeq$20\% for (2-1).}
\tablenotetext{b}{The displayed ratios are not corrected for differences 
in resolved out flux.  Asssuming the resolved out flux is constant over 
the map, the true ratios should be corrected up by a factor of 0.77/0.58 
= 1.33.}
\enddata
\end{deluxetable}

\clearpage

\begin{deluxetable}{ccccccc}
\tablenum{4}
\tablewidth{0pt}
\tablecaption{Giant Molecular Clouds in NGC 3077\label{tab4}}
\tablehead{
\colhead{GMC\tablenotemark{a}} 
&\colhead{$\alpha_{o}, ~\delta_{o}$\tablenotemark{e}}
&\colhead{$a \times b$} 
&\colhead{$\Delta v_{1/2}$}
&\colhead{$M_{vir}$}
&\colhead{$<n_{H_{2}}>$\tablenotemark{b}}
&\colhead{$M_{Xco}$\tablenotemark{c}}
\\
\colhead{} 
&\colhead{}
&\colhead{($pc \times pc$)} 
&\colhead{($km/s$)}
&\colhead{($10^{6}M_{\odot}$)}
&\colhead{($10^{3}cm^{-3}$)}
&\colhead{($10^{6}M_{\odot}$)}
}
\startdata
A &(20.24,03.2) &$77\times 19$ & 23 &2.7 &5.4 &1.0   \\
B &(19.08,00.1) &$68\times 23$ & 51 &14 &6.3 &2.3   \\
C &(18.88,51.3) &$<10\times <10$\tablenotemark{f} 
&23 &$<$0.70 &$>$7.6 &0.14   \\
D &(18.85,57.8) &$79\times 62$ &34 &11 &0.9 &1.0   \\
E &(16.3,08)& \nodata & \nodata & \nodata &\nodata &0.35    \\
F &(13.71,09.8)&$110\times 84$ & 12 & 1.8 &0.02 &1.1    \\
G &(12.6,17)& \nodata & \nodata & \nodata &\nodata &0.66   \\
\enddata
\tablenotetext{a}{Based on fits to the CO(2-1) data, except GMC B which 
is fit with the CO(1-0) data.}
\tablenotetext{b}{The average density is taken to be $M_{vir}
/m_{H2}Vf^{3/2}$ with $V = (4\pi/3)(0.7\sqrt{ab})^{3}$.  $f$ is assumed 
to be unity for GMC B.}
\tablenotetext{c}{Based on CO(1-0) with $X_{CO}=2.3\times 10^{20} cm^{-2}~
(K~km~s^{-1})^{-1}$, and corrected for resolved out flux.}
\tablenotetext{e}{Based on (10$^{h}$03$^{m}$;68$^{o}$44$^{'}$) for all 
GMCs except D and E, and (10$^{h}$03$^{m}$;68$^{o}$43$^{'}$) for D \& E.}
\tablenotetext{f}{Clouds are considered unresolved if the gaussian fit is 
smaller than 1/2 the beam minor axis ($\simeq$10 pc).}
\end{deluxetable}

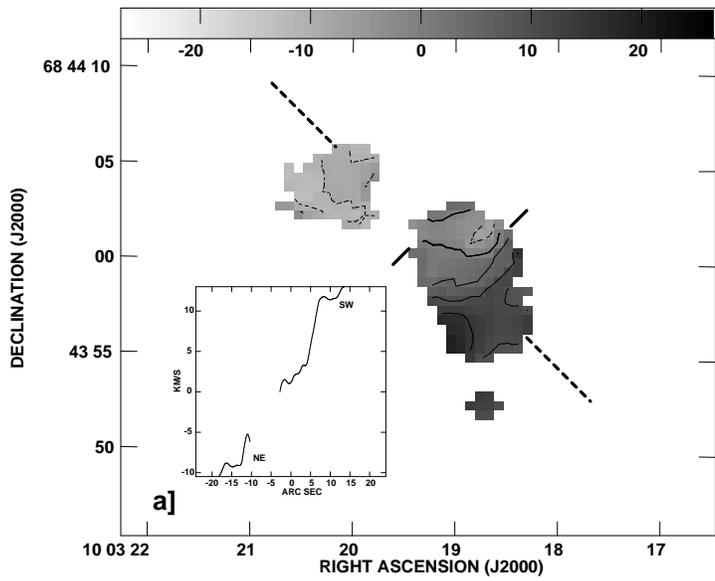
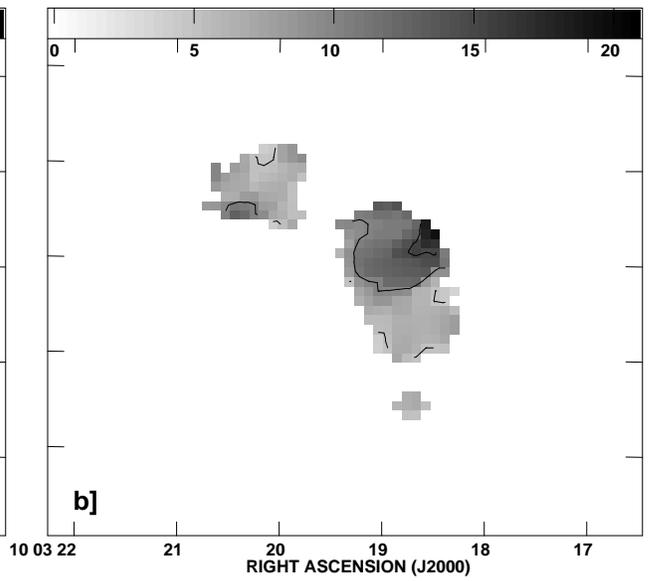
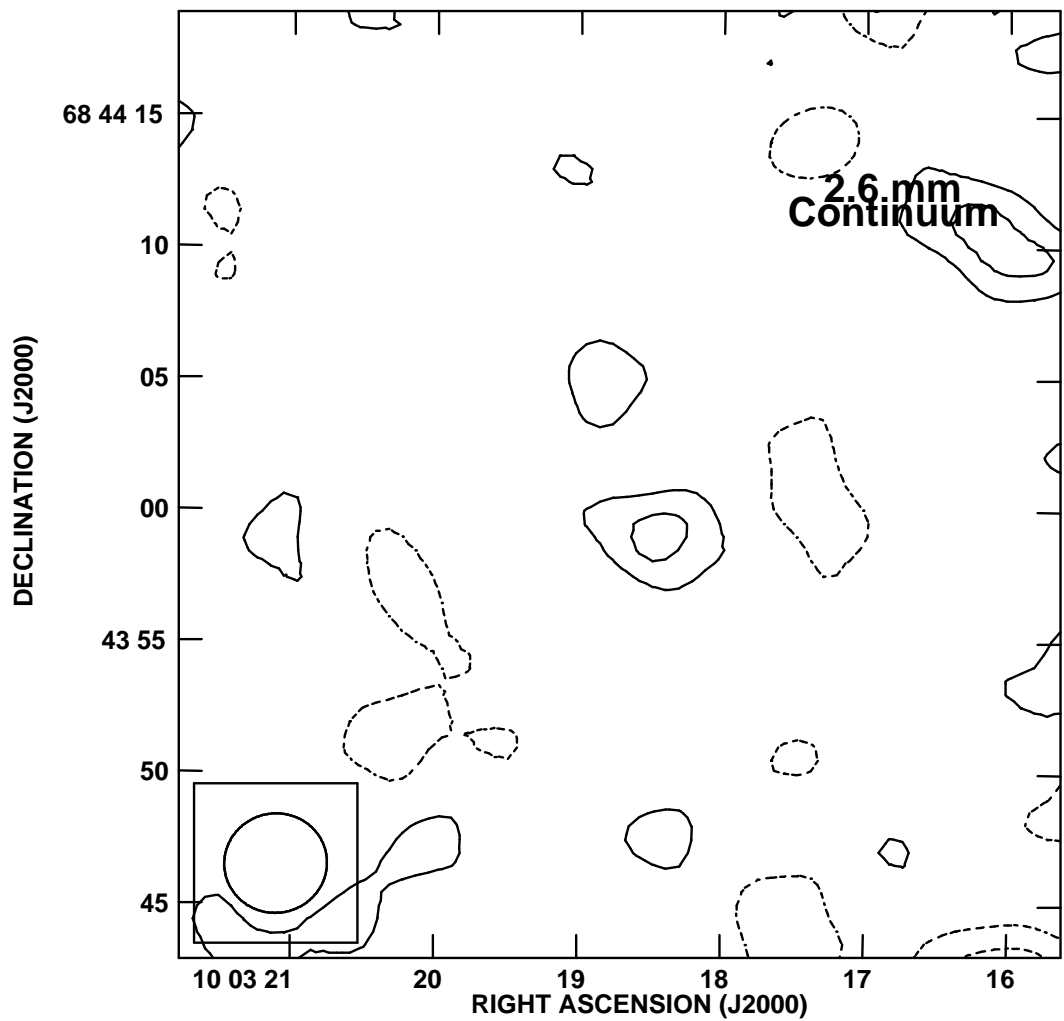
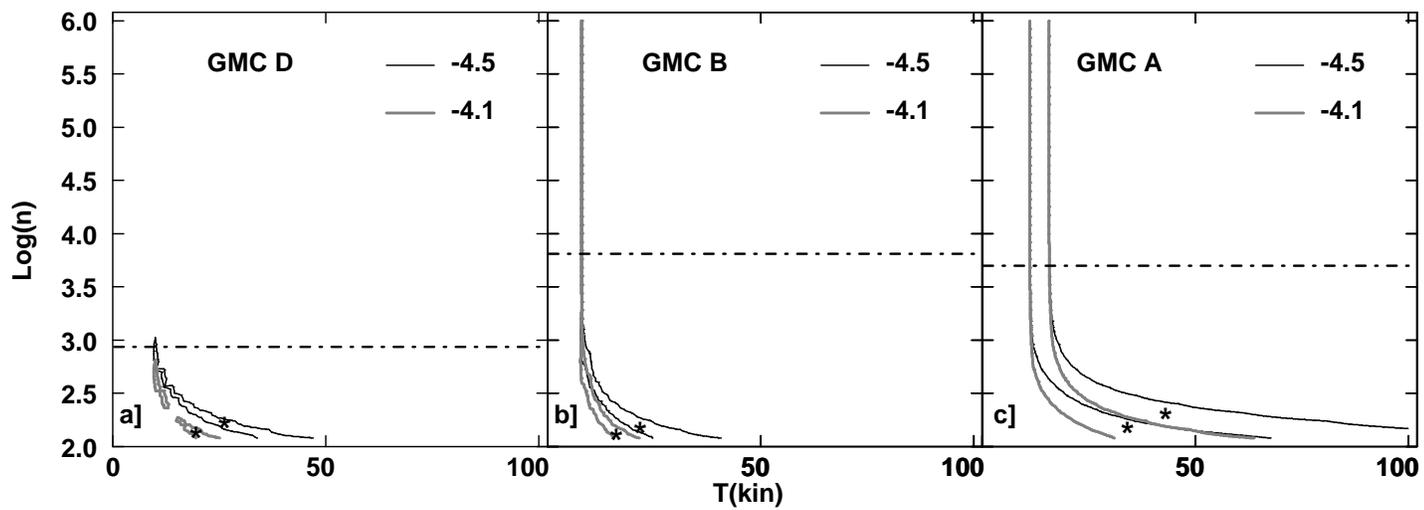
\end{document}